\documentclass[notitlepage,a4paper]{article}

\global\long\def\order#1{\mathcal{O}\left(#1\right)}
\global\long\def\su#1{{\mbox{\tiny #1}}}
\global\long\def\d{\mathrm{d}}
\def\Emax{E_{\max}}
\def\za{Z\alpha}
\usepackage{mathtools}
\usepackage{amsmath}
\usepackage{amssymb}
\usepackage{slashed}
\usepackage{graphicx}
\usepackage{epsfig}
\usepackage{epstopdf}
\usepackage{esint}
\usepackage[onehalfspacing]{setspace}
\usepackage[margin=1.0in]{geometry}
\begin{document}

\title{High-energy electrons from the muon decay in orbit: radiative corrections}

\author{Robert Szafron, Andrzej Czarnecki\\ \multicolumn{1}{p{.7\textwidth}}{\centering\emph{
{Department of Physics, University of Alberta, Edmonton,
  Alberta, Canada T6G 2G7}}}}

\maketitle

\begin{abstract}
 We determine the $\mathcal{O}(\alpha)$ correction to the energy
  spectrum of electrons produced in the decay of muons bound in
  atoms. We focus on the high-energy end of the spectrum that
  constitutes a background for the muon-electron conversion and will
  be precisely measured by the upcoming experiments Mu2e and COMET. The
  correction suppresses the background by about 15\%. 
\end{abstract}

%PACS 13.35.Bv \sep 36.10.Ee

\vspace{1cm}

In matter, muons decay differently from antimuons. Although the decay
rates are very similar \cite{Czarnecki:1999yj}, negatively charged
$\mu^-$ can bind with nuclei.  The nucleus exchanges photons with the
muon and the daughter electron, re\-arranging the energy distribution.
In this paper we find how this rearrangement is affected by the real
radiation and self-interaction on the muon-electron line. We predict
the energy spectrum of the highest-energy electrons, interesting both
theoretically and experimentally.

For a theorist, the muon decay is the simplest
example with which to understand the gamut of binding effects, including the
motion in the initial state, interplay of the binding and the
self-interaction, and the recoil of the nucleus. Experimenters have recently
studied the bound muon decay  (decay in orbit, DIO)
\cite{Grossheim:2009aa} with a precision sufficient to probe radiative
corrections, later evaluated in \cite{Czarnecki:2014cxa};
however, these studies concern only the lower half of the spectrum, largely
accessible also to a free muon.

Interestingly, the energy range of electrons produced in the DIO reaches to about twice the
maximum possible in a free muon decay. When the muon decays in vacuum,
momentum conservation requires that at least half of the energy be
carried away by the neutrinos. In the DIO, the nucleus can absorb the
momentum without taking much energy because it is so heavy.

The high-energy part is important for the upcoming searches for the
ultra-rare neutrinoless muon-electron conversion, COMET in J-PARC
\cite{Kuno:2013mha} and Mu2e in Fermilab \cite{Brown:2012zzd}.
Designed for a sensitivity better than one exotic conversion in
$10^{16}$ ordinary muon decays, they will collect large samples of
events with high-energy electrons.  A reliably predicted spectrum is
needed to distinguish the exotic signal -- an excess of electrons at
maximum energy -- from the Standard Model background.

Predicting the DIO spectrum is a challenge because both the decaying
muon and the daughter electron interact with the Coulomb field of the
nucleus. A numerical calculation with Coulomb-Dirac wave functions is
possible \cite{Czarnecki:2011mx} provided that self-interactions
(photons attached to the muon and the electron) are neglected. How can
they be included? In the lower half of the spectrum the muon and the
electron can be treated as nearly free and the binding effects can
be factorized.  Then the radiative corrections, known for a free
muon, are convoluted with a shape function that parametrizes the
Coulomb field effect \cite{Czarnecki:2014cxa,Szafron:2015mxa}.
Here we construct an expansion around the
end-point and employ it to  find radiative corrections also to the high-energy part of the spectrum.

\begin{figure}[htb]
    \centering
    \includegraphics[width=0.45\textwidth]{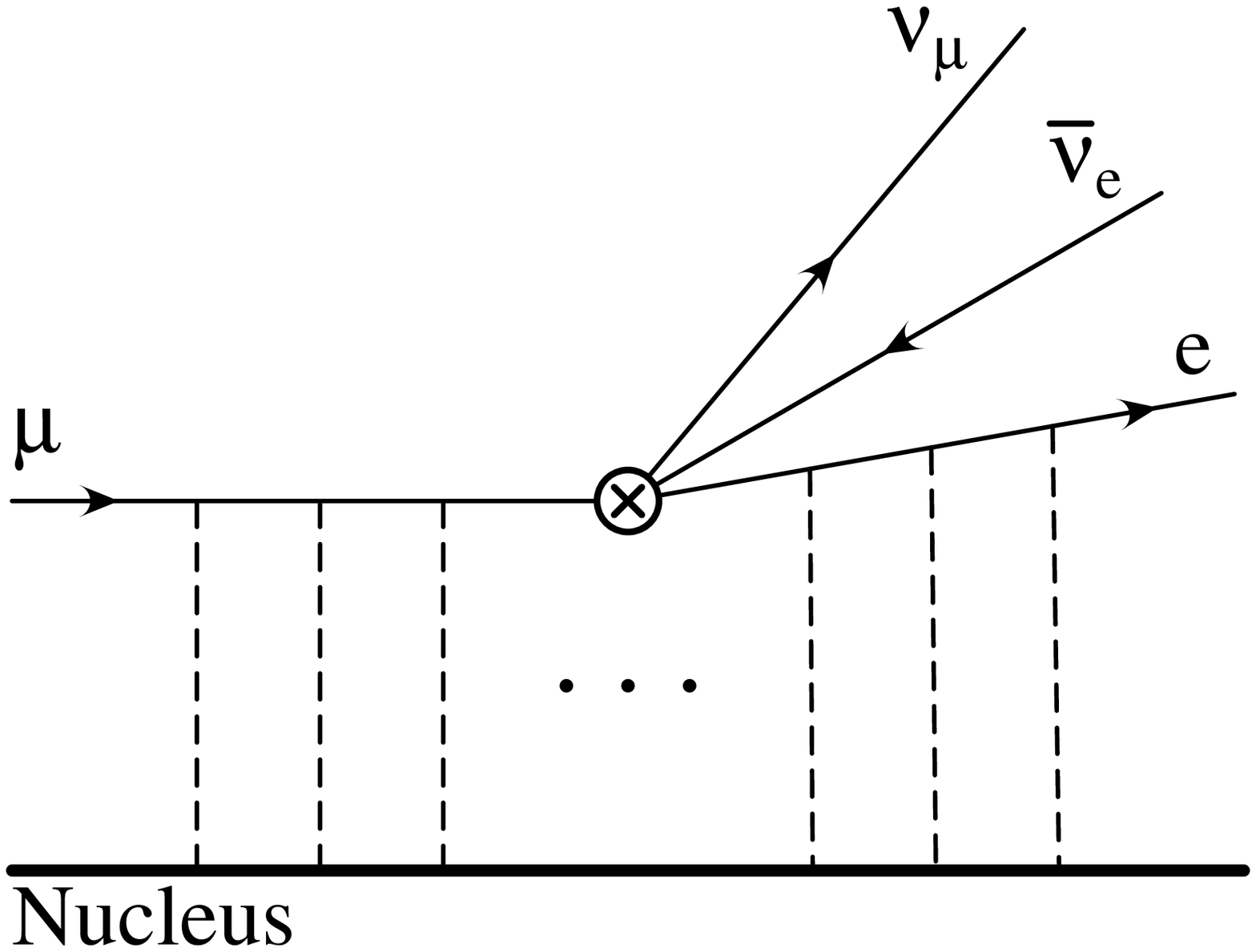}\hfill
    \includegraphics[width=0.45\textwidth, trim=0 -3cm 0
    -2cm]{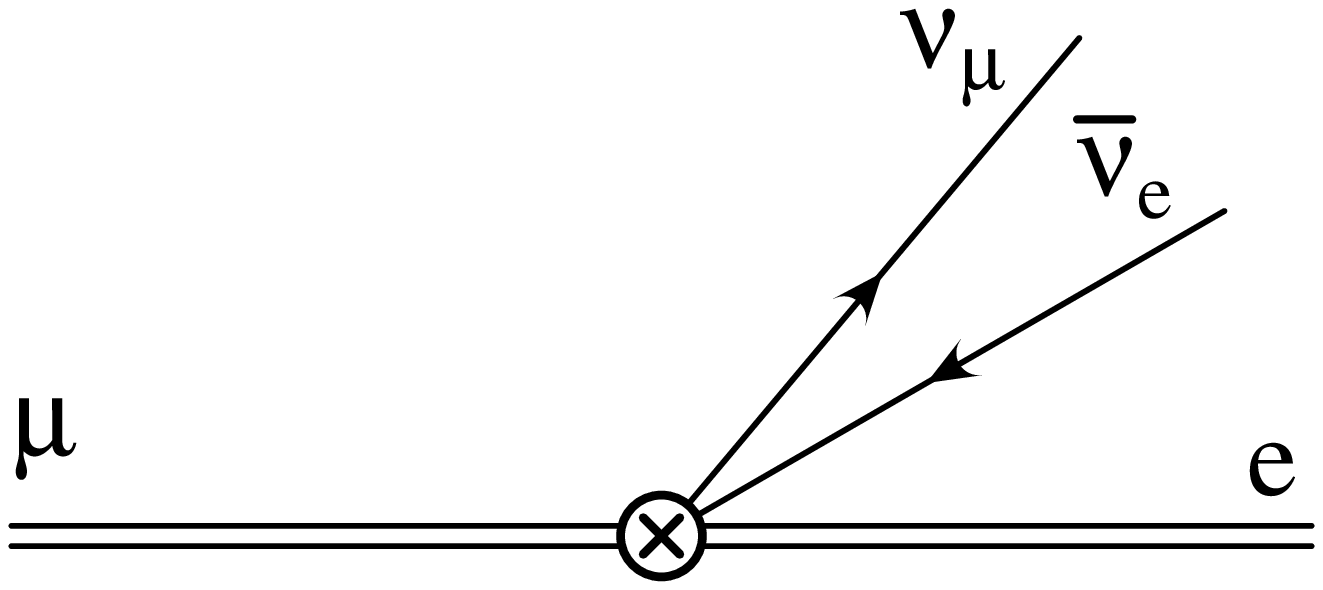}\\[-7mm]
\hspace*{50mm}\vspace*{-10mm}Furry picture\vspace*{10mm}
    \caption{Muon decay in orbit (DIO). Dashed lines denote
      Coulomb photons exchanged between charged leptons and the 
      nucleus. The right panel shows the same physics using 
      double lines for charged leptons propagating in the 
      Coulomb field.}
    \label{fig:diag1}
\end{figure}

Accounting for the external Coulomb field in charged-particle
propagators is called the Furry picture \cite{Furry:1951zz}.  In this
formulation, and still ignoring radiative corrections, a single
diagram, shown in Fig.~\ref{fig:diag1}, describes the DIO.
We shall demonstrate that the bound-state radiative corrections are
easiest to evaluate near the high-energy end of the spectrum,
the most important part for the new experiments. For now we neglect the
nuclear recoil and structure, and treat the nucleus as an
infinitely-heavy point source of the Coulomb field.  We denote the
electron energy with $E$; its maximum value is $\Emax \simeq m_\mu\left(1  - \frac{(\za)^2}{2} \right)$,
where $m_\mu$ is muon mass, $Z$ is the atomic number, and
$\alpha\simeq 1/137$ is the fine-structure constant. The DIO spectrum  near its end-point
can be expanded in the small parameter
$\Delta = \frac{\Emax-E}{m_\mu}$,
\begin{equation}\label{eq:expand}
\frac{m_\mu}{\Gamma_0}\frac{\d\Gamma}{\d E} = \sum_{ijk} B_{ijk}\Delta^i
(\pi \za)^j \left( \frac{\alpha}{\pi} \right)^k,
\end{equation}
where $\Gamma_0 = \frac{G_\su{F}^2 m_\mu^5  }{192\pi^3}$ is the free-muon
decay rate and $G_\su{F}$ is the Fermi constant \cite{Webber:2010zf,Marciano:1999ih}.
Powers of $\pi \za $ parameterize photon exchanges with the nucleus
and $\alpha/\pi$ arises from radiative corrections on the
charged-lepton line and the vacuum polarization.  The first
non-vanishing term has $i=j=5$ and $k=0$, with
$B_{550}=\frac{1024}{5\pi^6 }\simeq 0.21$. 
Higher order coefficients $B$ may have logarithms of $\za$
and $\Delta$. 

Corrections to this leading behavior have several sources. The large momentum transfer to
the nucleus probes its interior. The finite nuclear size, already included in
\cite{Czarnecki:2011mx}, causes the largest correction.  We will
comment at the end of this paper on how to include it in our
formalism. The finite nuclear mass introduces a recoil effect, also
evaluated in \cite{Czarnecki:2011mx}. It affects the coefficients $B$
only slightly but it shifts the end-point energy $\Emax$.
 
We shall exploit a theoretical similarity between the DIO and the
photoelectric effect to control higher-order binding effects. They
generate powers of $\pi Z \alpha$ \cite{Pratt:1973na,Eichler20071}
rather than $Z\alpha$.  Indeed, a numerical evaluation for a point
nucleus with $Z=13$ (as in aluminum, the planned target in COMET and
Mu2e) finds a $-21\%$ correction, consistent with
$13\pi \alpha = 0.3$.  Logarithmic enhancement starts with
$(\pi Z\alpha)^7\ln(Z\alpha)$.  Fortunately, these large effects,
slightly suppressed by the finite nucleus size, are summed up in the
numerical evaluation \cite{Czarnecki:2011mx}.

Finally, the most challenging corrections result from radiative
effects that are the subject of this study. Before delving into the physics
of the end-point, we present our main result. Close to the end-point,
including radiative corrections, the DIO spectrum for aluminum is
\begin{equation}
\frac{m_\mu}{\Gamma_0}\frac{\d\Gamma}{\d E} 
\approx 1.24(3) \times 10^{-4}\times \Delta^{5.023} .
\label{eq:f}\end{equation}
To illustrate the importance of the new corrections we consider the
last 150 keV of the spectrum (the typical planned resolution of Mu2e and COMET). Radiative
corrections reduce the number of events in this bin by $15\%$, a welcome
reduction of the  background,  comparable in size with
higher-order binding effects. 

In the remainder we explain the origin of such a large effect. We
begin with the tree-level behaviour, appropriately expanding the
lepton wave functions. We find that an exchange of a single, highly
virtual photon gives the electron an energy  of the full muon mass.

The relativistic electron is described by a  plane wave distorted by
the Coulomb potential $V$; to the first order,
\begin{equation}
\label{eq:WFe}
\overline{\psi}_p\left(\vec q\right) =
\bar{u}(p)\left[ \delta^{3}\left(\vec{p}-\vec{q}\right) 
+\slashed V\left((\vec{p}-\vec{q})^2\right)   \frac{1}{\slashed{q}-m_e }\right],
\end{equation}
where $u(p)$ is a spinor solution of a free Dirac equation and the
four-potential in  momentum space reads
\begin{equation}\label{eq:point}
V\left(\vec{k}^2\right)=\left(-\frac{ Z\alpha}{2\pi^2\vec{k}^2},\vec{0}\right).
\end{equation}  
A muon bound to a nucleus with $Z\ll 137$ is
nonrelativistic. Nevertheless, we will need the first relativistic
correction to its wave function, just like in the classic analysis of
the photoelectric effect \cite{Berestetsky:1982aq},
\begin{eqnarray}\label{eq:WFm}
\psi\left(\vec q\right) &=& \psi_\su{NR}\left(\vec q\right)
\left(1+\frac{\vec{q}\cdot \vec{\gamma}}{2m_\mu}\right)u(P),
\end{eqnarray}
where
$\psi_{\su{NR}}\left(\vec q\right) = \frac{8\pi Z\alpha m_\mu \Psi(0)
}{\left[\vec{q}^2+(Z\alpha m_\mu)^2\right]^2}$
is the nonrelativistic momentum-space wave function of the 1S ground
state with
$\Psi(0)=\left(\frac{Z\alpha m_\mu}{\pi^{1/3}} \right)^{3/2}$; $u(P)$
is the four-spinor of a muon at rest, $P=(m_\mu,0)$.

\begin{figure}[ht]
    \centering
    \includegraphics[width=0.45\textwidth, trim = 0 0 0 -4mm]{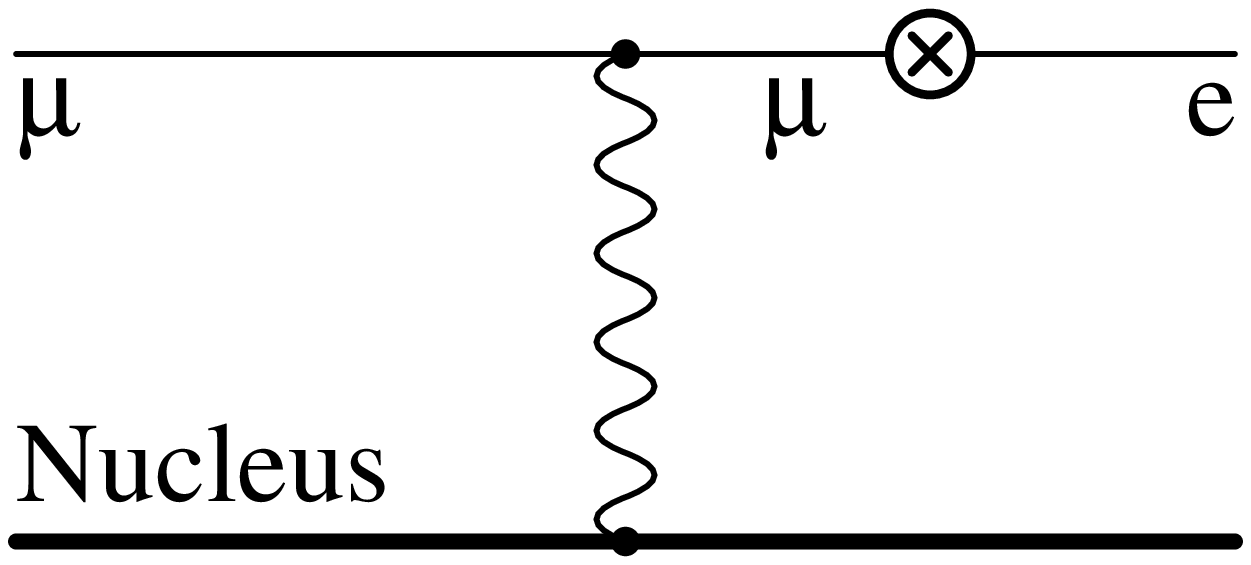}
\hspace*{3mm}
    \includegraphics[width=0.45\textwidth, trim = 0 0 0 -4mm]{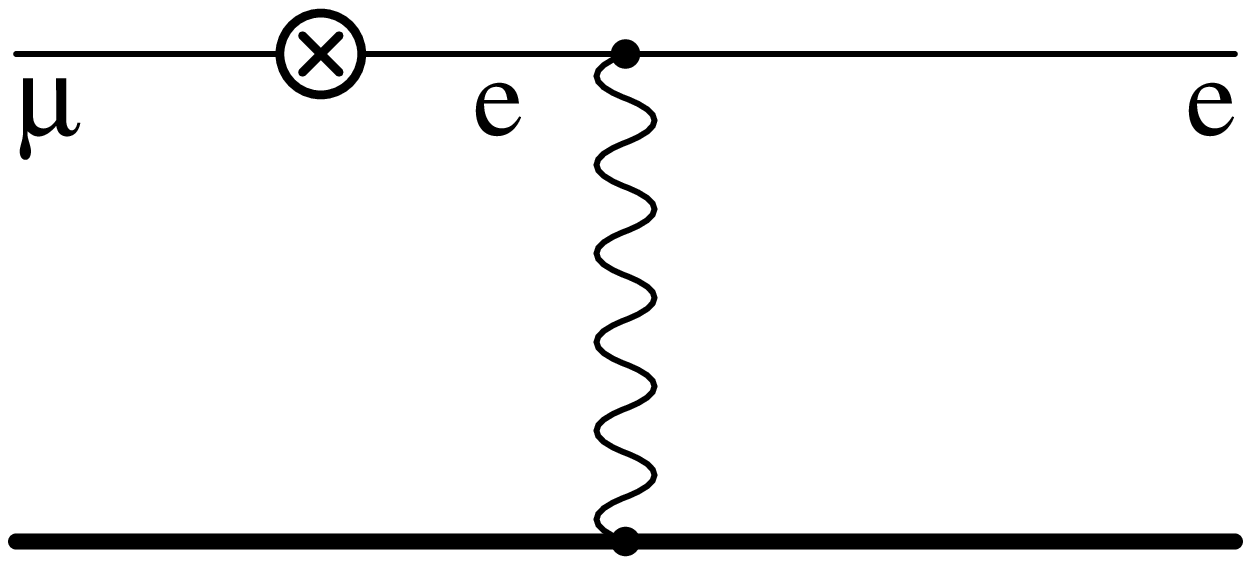}
\\
 (a) \hspace*{38mm} (b)
 \caption{Furry diagram expanded in $Z\alpha$.  Crossed circles
   indicate insertions of the weak interaction transforming the muon
   into an electron; the emitted neutrinos are not shown. These two
   amplitudes give rise to the highest-energy electrons.}
    \label{fig:diag2}
\end{figure}

We now consider separately the contributions of the two terms in the
electron wave function \eqref{eq:WFe}. The delta function term forces
the muon momentum in \eqref{eq:WFm} to be large, $\vec q =\vec p \sim
m_\mu$. Thus we neglect $Z\alpha m_\mu$ in the denominator of $\psi_{\su{NR}}$ and find
\begin{equation}\label{eq:WFm1}
\psi\left(\vec q\right) \approx  (2\pi)^3\Psi(0)\frac{1}{\slashed P +
  \slashed q-m_\mu}\slashed V\left({\vec{q}}^{\, 2}\right) u(P).
\end{equation}
This is visualized in Fig.~\ref{fig:diag2}a: the muon, before decaying, transfers
momentum $\vec q \sim m_\mu$ to the nucleus through a hard
space-like photon.  It is here that the relativistic correction to the
muon wave function is important. 

The second term in \eqref{eq:WFe} refers to an electron scattered on
the nucleus. Now the muon momentum, not restricted to large values,
has its typical bound-state size $\vec q \sim Z\alpha m_\mu$,
negligible in comparison with $\vec p \sim m_\mu$. We use
$\lim_{a\to 0} \frac{8\pi a}{(q^2 + a^2)^2}= (2\pi)^3
\delta^3\left(\vec q\right)$ to approximate the muon wave function,
\begin{equation}\label{eq:WFm2}
\psi\left(\vec q\right) \approx  (2\pi)^3\Psi(0) \delta^3\left(\vec
  q\right) u(P).
\end{equation}
This is shown in Fig.~\ref{fig:diag2}b, where the hard photon is
exchanged after the decay. 

The two diagrams in Fig.~\ref{fig:diag2} add up to the leading
contribution $B_{550}$ in \eqref{eq:expand}. In both cases any energy
unused by the electron ($\sim \Delta$) is taken up by the neutrinos
and not transferred to the nucleus. Counting neutrino momenta in the
integrated matrix element explains the leading energy dependence in
\eqref{eq:expand},
\begin{equation}
  \label{eq:1}
  \int\frac{\d^3 \nu}{\nu_0} \frac{\d^3 \overline\nu_0}{\overline\nu_0}
 \delta\left( m_\mu\Delta - \nu_0- \overline\nu_0 \right)\ldots 
 {\slashed\nu}  \ldots\overline  {\slashed\nu} \sim \Delta^5.
\end{equation}

\begin{figure}[ht]
    \centering
    \includegraphics[width=0.45\textwidth]{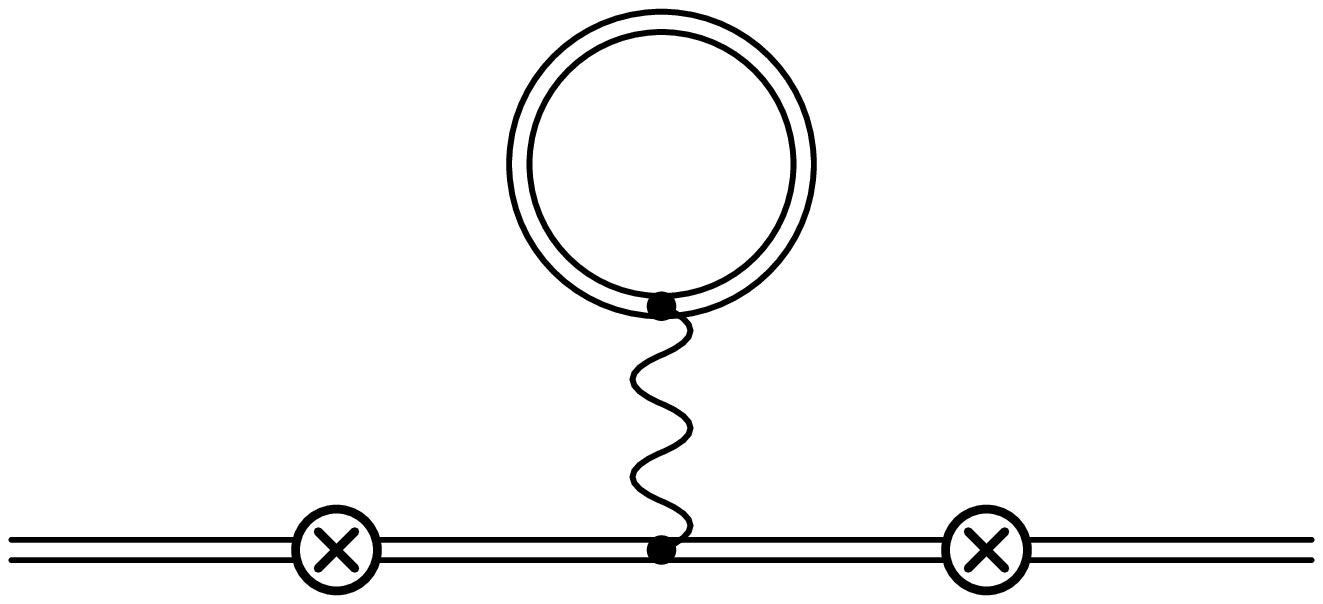}
\hfill 
    \includegraphics[width=0.45\textwidth]{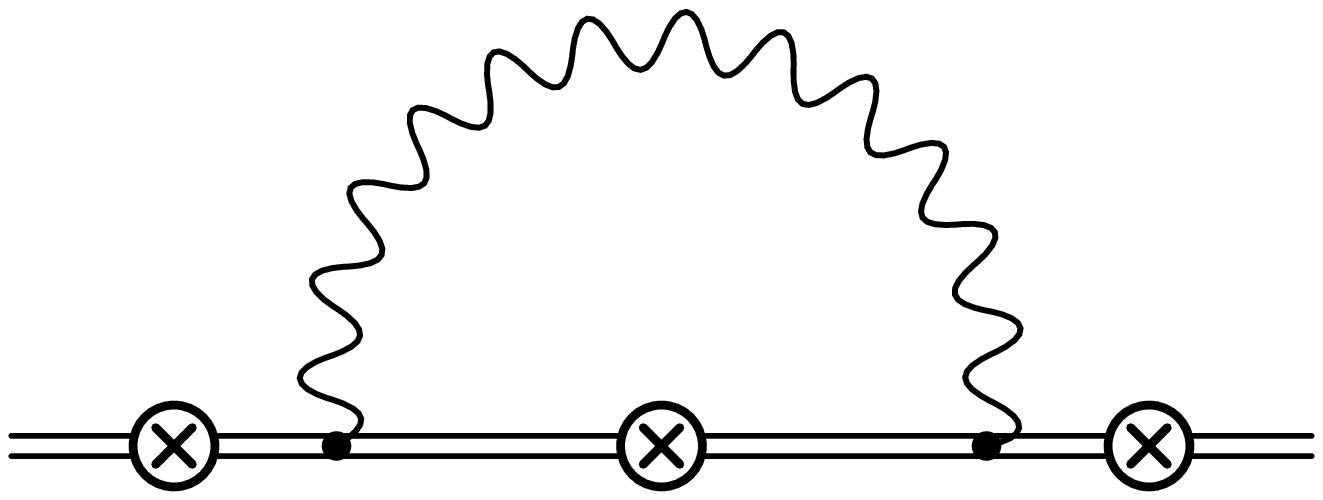}
\\
 (a) \hspace*{41mm} (b)
    \caption{Virtual corrections to the muon DIO (Furry
      picture).}
    \label{fig:diag3}
\end{figure}

Having understood that only two diagrams describe the end-point
behavior, we are now ready to evaluate radiative corrections.  In the
Furry picture there are two groups of virtual corrections, shown in
Fig.~\ref{fig:diag3}. We expand them in $\za$ just like the tree-level
diagrams, but in addition to wave functions (\ref{eq:WFe},~\ref{eq:WFm}), we need also the Coulomb-Dirac Green's function
\cite{Schwinger:1989ka},
\begin{eqnarray}\label{eq:GF}
-iG^V\left(E;\vec{p},\vec{p'}\right)&\simeq &
\frac{\delta^3\left(\vec{p}-\vec{p'}\right)}{\slashed p-m}
\nonumber \\
&& 
+\frac{1}{\slashed p-m} \slashed V\left((\vec{p}-\vec{p'})^2\right)
  \frac{1}{\slashed  p'-m}.
\end{eqnarray}
The expansion (\ref{eq:GF}) reduces radiative corrections in an
external field to a set of loop diagrams that we evaluate analytically
\cite{Smirnov:2014hma}.  This approach can be extended to higher-order
corrections.

The diagram in Fig.~\ref{fig:diag3}a adds the Uehling correction to
the Coulomb potential \cite{Uehling:1935uj} and modifies the photon
propagators in Fig.~\ref{fig:diag2}.  In muonic aluminum the range of
the Uehling potential exceeds the Bohr radius. It strengthens the
attractive force and increases the muon wave function at the origin,
\begin{equation}
\Psi(0) \rightarrow \Psi(0)\left(1+\frac{\alpha}{\pi}\delta_0\right).
\label{eq:psi0}
\end{equation}
For aluminum we find $\delta_0=3.27$. This correction reflects the
running of the coupling $\alpha$ up to the average muon momentum scale
$m_\mu Z\alpha$.

Vacuum polarization loops on the highly virtual photon propagators are
related to the running of $\alpha$ up to the hard scale $m_\mu$. They
enhance the tree-level decay rate by a factor
$1+\frac{\alpha}{\pi}\delta_\su{VP}$, with
\begin{equation}
\delta_\su{VP}=\frac{4}{3}\ln\frac{m_\mu}{m_{e}}-\frac{10}{9}+0.12
 \approx 6.1,
\label{eq:vpwithmu}
\end{equation}
where the term 0.12 arises from a muon loop.

\begin{figure}[ht]
    \centering
    \includegraphics[width=0.45\textwidth]{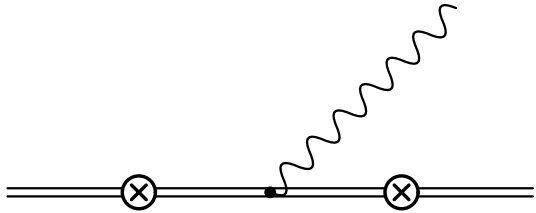}
    \caption{Furry diagram for the real radiation correction.}
    \label{fig:diag6}
\end{figure}

Another correction comes from the real radiation. Diagrams
represented by  Fig.~\ref{fig:diag6}
are expanded in the same way as virtual corrections, using
(\ref{eq:GF}). Near the end-point the eikonal approximation suffices;
by energy conservation the real photons must be soft,
$0 <E_\gamma <m_\mu \Delta$.

The sum of virtual and real  radiation is   finite,
\begin{equation}
  \label{eq:2}
\frac{B_{551}}{B_{550}}= \delta_\su{H} +\delta_\su{S} \ln\Delta,
\end{equation}
where the vacuum polarization corrections $\delta_\su{VP}$ and
$\delta_0$ are included with hard self-interaction effects in
$\delta_\su{H} =6.31 -\frac{26}{15}\ln\frac{m_{\mu}}{m_{e}} $, and
$\delta_\su{S}=2\ln \frac{2m_{\mu}}{m_{e}}-2$ is a soft correction.
The latter can be exponentiated \cite{Yennie:1961ad} (similarly to the
free-muon decay \cite{Marciano:1975kw}) and vanishes when
$\Delta \rightarrow 0$,
\begin{equation}\label{eq:radcor}
B_{550} + \frac{\alpha}{\pi} B_{551}
\to
B_{550}\left( \Delta^{\frac{\alpha}{\pi}
    \delta_\su{S}}+\frac{\alpha}{\pi} \delta_\su{H} \right),
\end{equation}
instead of unphysically diverging as $\ln\Delta$.  It increases the
exponent of $\Delta$ and suppresses DIO events near the end-point. The
relative decrease is inversely correlated with the energy resolution:
the number of electrons in the end-point bin of 1 (0.1) MeV is reduced
by 11\% (16\%).

The final-state electron is relativistic, $E \gg m_e$, so its
structure function \cite{Arbuzov:2002pp,Arbuzov:2002cn} is insensitive
to Coulomb corrections. A convolution with the free-decay spectrum
confirms the coefficient $-\frac{46\alpha}{15\pi}$ of
$\ln\frac{m_\mu}{m_e}$ due to collinear photons. Together with the
vacuum polarization in (\ref{eq:vpwithmu}), this explains the
logarithmic part of the hard correction.

That log is largely cancelled in the sum with the wave function
correction in (\ref{eq:psi0}) and $\delta_\su{H} = -2.9$ reduces the
end-region by only a fraction of a per cent.  We thus neglect the
unknown hard corrections $\order{(\alpha/\pi)^2}$ in the error
estimate, dominated by the nuclear-size effects, discussed below.

There are now two complementary studies of the end-point
spectrum. Here, we have computed radiative corrections (RC) assuming a
point nucleus and considering only the one-Coulomb
exchange. Ref.~\cite{Czarnecki:2011mx} did not have the RC but
included the nucleus structure, recoil, and multiple Coulomb
interactions.

In order to combine these results, we observe that the most important
-- soft -- correction is universal, not sensitive to any interactions
with the nucleus. The hard correction is tiny, so treating it also as
universal is well within our final error estimate. 

In the discussion of the uncertainty we specialize to
aluminum but the discussion can be applied to other nuclei, so we keep
the $Z$ dependence explicit.  We assume a Fermi charge distribution,
\begin{equation}\label{eq:charge}
\varrho=\frac{\rho_0}{1+\exp\frac{r-r_{0}}{a_0} }
\end{equation}
with $a_0=0.569 $ fm and $r_0=2.84(5) $ fm \cite{vries87}.

The finite size affects the nucleus form-factor, defined as a ratio of Fourier
transforms of potentials from the extended (\ref{eq:charge}) and the
point-like (\ref{eq:point}) charge distributions,
\begin{equation}
F_\rho(\vec{k}^2)=\frac{V_\rho\left(\vec{k}^2\right)}{V\left(\vec{k}^2\right)}\longrightarrow
0.64 \;\text{ for } \; \vec{k}^2=m_\mu^2.
\end{equation} 
The DIO spectrum for a finite nucleus has an expansion
analogous to \eqref{eq:expand}, but with coefficients 
that depend on the density $\rho$. Its leading term near the end-point 
\cite{Czarnecki:2011mx} is 
\begin{equation}\label{eq:sum}
\sum_{j=5}^{\infty}B_{5j0}^{\rho}\left[F_\rho\left(m_\mu^2\right)  \pi  Z\alpha\right]^j
=8.98\times 10^{-17}\left( \frac{m_\mu}{\text{MeV}} \right)^6.
\end{equation} 
This result includes exchanges of many Coulomb photons, in addition to
the single hard exchange to which we have found the radiative
correction. We estimate the magnitude of the multi-Coulomb part as a
fraction $f=F_\rho\left(m_\mu^2\right)  \pi  Z\alpha \simeq 0.2$ of
(\ref{eq:sum}).

Hard radiative corrections to this part are missing.  To be
conservative, we are not assuming that they involve a cancellation
that has suppressed $\delta_\su{H}$.  Corrections on the order of the
collinear logarithm translate into a relative error of about
$\frac{46\alpha}{15\pi}f \ln\frac{m_\mu}{m_e} \simeq 0.7\%$. In
addition, experimental errors in the charge distribution parameters
(\ref{eq:charge}) introduce a 2\% uncertainty
\cite{Czarnecki:2011mx}. Summing them in quadrature, together with the
sensitivity to the scale involved in the exponentiation of soft
effects, we arrive at an error around 2.5\% in the end-point spectrum.

The result (\ref{eq:sum}), multiplied by the new correction
\eqref{eq:radcor}, leads to our prediction for the end-point spectrum,
(\ref{eq:f}).

To summarize, we have determined the correction to the high-energy
tail of the DIO energy distribution and its remaining uncertainty. Key
to this improvement has been the simplicity of the leading amplitudes
that turn out to arise from a small number of hard-photon
exchanges. This line of reasoning can be extended to higher-order
binding effects, at least for a point nucleus. For a realistic 
charge distribution, a numerical evaluation of loop diagrams will be
necessary. However, the leading radiative correction has now been
established with good precision.  Its sizeable negative effect on the
DIO will make any observed event near the end-point an even more
convincing signal of New Physics, a discovery we eagerly anticipate.

%\paragraph*{}

This research was supported by Natural Sciences and Engineering
Research Council (NSERC) of Canada.  R.S.~acknowledges  support
by the Polish National Science Centre (NCN) under Grant Agreement
No.~DEC-2013/11/B/ST2/04023 and by the Fermilab Intensity Frontier
Fellowship. Fermilab is operated by Fermi Research Alliance, LLC under
Contract No.~De-AC02-07CH11359 with the United States Department of
Energy.

\end{document}